\documentclass[conference]{IEEEtran}
\IEEEoverridecommandlockouts
\usepackage{cite}
\usepackage{amsmath,amssymb,amsfonts}
\usepackage{algorithmic}
\usepackage{graphicx}
\usepackage{tabularx}
\usepackage{textcomp}
\usepackage{xcolor}
\usepackage{multirow}
\usepackage{caption}
\usepackage{amssymb}
\def\BibTeX{{\rm B\kern-.05em{\sc i\kern-.025em b}\kern-.08em
    T\kern-.1667em\lower.7ex\hbox{E}\kern-.125emX}}

\begin{document}

\title{Acoustic UAV Detection in Battlefield Scenarios: Handling Noise, Domain Shift, and Weak Labels}
% MicShift: Mitigating Device Shift for Acoustic UAV Detection
% MADDNS: Mitigating Acoustic Device Domain Shift
% SHUM: Shift by Heterogeneous UAV Microphones and its mitigation
% Mitigation of SHUM: data/domain Shift by Heterogeneous UAV Microphones
% Practical Acoustic UAV Detection: Addressing Data Scarcity and Sensor Variability with Ukrainian Battlefield Data
% Acoustic UAV Detection in Battlefield Scenarios: Handling Noise, Domain Shift, and Weak Labels
% Mitigating Environmental Interference and Sensor Domain Shift in Acoustic UAV Detection

\author{
\IEEEauthorblockN{
Vadym Vilhurin\textsuperscript{1,2},
Volodymyr Sydorskyi\textsuperscript{1,2,3},
Andrii Shevtsov\textsuperscript{2,3},
}
\IEEEauthorblockA{
\textsuperscript{1}Institute for Applied System Analysis, Igor Sikorsky Kyiv Polytechnic Institute, Kyiv, Ukraine
}
\IEEEauthorblockA{
\textsuperscript{2}Zvook, Lviv, Ukraine
}
\IEEEauthorblockA{
\textsuperscript{3}Computer Science Department, Kyiv School of Economics, Kyiv, Ukraine
}
Emails: vilhurin.vadym@lll.kpi.ua, volodymyr.sydorskyi@gmail.com, ashevtsov@zvook.tech
}

\maketitle

\begin{abstract}
Passive acoustic sensing offers a critical, cost-efficient, and, crucially, passive alternative for detecting small unmanned aerial vehicles. However, the practical deployment of acoustic systems is discouraged by extreme environmental noise and sensor-induced domain shift caused by heterogeneous hardware. This paper addresses these challenges by introducing a robust framework optimized for real-world battlefield conditions. We propose the integration of Per-Channel Energy Normalization (PCEN) and attention-based pooling to enhance feature extraction under low signal-to-noise ratio scenarios. We further propose a domain-aware training strategy that leverages auxiliary classes and multi-microphone data to mitigate cross-domain performance degradation. Evaluated on a unique dataset of combat-zone recordings from the Ukrainian frontlines, our approach significantly outperforms existing baselines, increasing the F1 score from 55.4\% to 78.6\%.

This paper was originally presented at the International Conference on Military Communication and Information Systems (ICMCIS), organized by the Information Systems Technology (IST) Scientific and Technical Committee, IST-224-RSY – the ICMCIS, held in Bath, United Kingdom, 12-13 May 2026.
\end{abstract}


\section{Introduction}
The rapid development and mass deployment of small unmanned aerial vehicles (UAVs) pose a growing challenge for reliable detection in military and security-critical environments. In battlefield and contested-area scenarios, UAVs represent a direct threat to critical infrastructure, personnel safety, and operational secrecy. Early and robust detection under realistic conditions characterized by strong environmental interference, clutter, and heterogeneous sensing hardware is essential for estimating UAV trajectories, identifying operational zones, and enabling timely countermeasures \cite{Dong2025SecuringTS, Zhu2025IntelligentP}.
 
A wide range of approaches have been explored for UAV detection, including vision-based methods using RGB or thermal imagery \cite{Kaur2025Analysis, Kurmashev2025StudyOT, Adeeba2025VisionAD}, radio-frequency (RF) sensing \cite{Shi2025SmallUAV, Zhu2025IntelligentP}, and radar- or infrared-based systems \cite{Aziz2023AnalysisO}. While effective in controlled settings, these techniques face significant limitations in real operational environments. Vision-based detection requires clear line-of-sight, favorable lighting, large storage and computational resources, which complicates large-scale deployment. RF-based methods depend on prior knowledge of communication protocols and are vulnerable to interference from other RF-emitting devices common in battlefield settings. Infrared and radar systems require specialized, costly hardware and may be constrained by environmental or operational factors. 
In contrast, acoustic UAV detection is passive, low-cost, protocol-agnostic, and largely invariant to visual occlusion, making it particularly attractive for early detection in cluttered and contested environments. These properties are especially valuable in battlefield scenarios, where stealth, rapid deployment, and robustness to partial occlusion are critical \cite{Zhu2025IntelligentP, Shi2025SmallUAV}. 

Despite these advantages, practical acoustic UAV detection remains challenging. Real-world environments contain strong and highly variable background noise from wind, vehicles, machinery, and human activity, which can obscure UAV signatures and significantly degrade detection performance \cite{Dong2025SecuringTS}. Moreover, acoustic data collected from different microphones and deployment locations often exhibit substantial domain shift, driven by variations in sensor frequency response, placement, and environmental conditions \cite{AlEmadi2021AudioBasedDD}. These challenges are further compounded by severe data scarcity in the operational domain, where only limited labeled recordings are typically available.

Existing work addresses environmental interference, cross-microphone generalization, and data scarcity jointly. This paper focuses on this combined setting, investigating acoustic UAV detection under realistic noise conditions and strong sensor-induced domain shift.
The contributions of this work include:
\begin{itemize}
    \item We introduce a novel challenge of adapting a model trained across two acoustic domains that differ in recording device configuration, sensor location, typical background noise, and target classes.
    \item We propose a robust Convolutional Neural Network (CNN)-based acoustic classification framework that combines Per-Channel Energy Normalized (PCEN) spectrograms with an attention-based pooling mechanism to improve robustness to environmental noise and weakly-labeled data
    \item We introduce a domain-aware training strategy that leverages data from multiple microphone domains and employs noise-driven and curriculum-based augmentations to mitigate cross-domain shift between microphone domains.
    \item We show that including auxiliary UAV classes from the auxiliary domain in the training dataset improves generalization to the target UAV class in the target domain.
\end{itemize}

All the data used for this research are provided by Zvook \cite{zvookwebsite} and collected from Ukrainian frontlines, training polygons, and active anti-aircraft acoustic sensors. This increases the practical value of this research.

\section{Rationale to Military Application}

Acoustic UAV detectors offer effective and efficient solutions to the modern small UAV sensing problem. As passive systems, they are far less susceptible to radio jamming or electronic countermeasures, enhancing their reliability and stealth in contested environments. This makes them particularly valuable in real warfare scenarios of drone-centric Russia-Ukraine war, or NATO's border surveillance, where they can integrate into a multi-layered defense against drone swarms. Their low cost and ease of mass production further enable deployment in large-scale battlefield operations. Additionally, acoustic sensors deliver high performance and accuracy at short- to medium-range distances.

In acoustic detection, domain shift mitigation is essential for maintaining operational effectiveness in modern warfare. As environmental conditions fluctuate with seasons and new UAV models introduce novel acoustic signatures, gathering vast datasets for every possible scenario becomes resource-intensive and impractical \cite{AlEmadi2021AudioBasedDD}. Unaddressed domain shifts could result in critical failures, such as missed detections or false alarms, especially under adversarial tactics like sound mimicry. Therefore, any machine learning-based military system must incorporate adaptability to changing conditions from the outset, ensuring resilience in dynamic and high-stakes theaters.

\section{Related Work}

Acoustic UAV detection typically relies on identifying mechanical signatures such as blade-pass frequency harmonics and amplitude modulations, which form distinct patterns. Early work demonstrated that spectral features combined with Support Vector Machine can separate drones from background noise under controlled conditions \cite{bernardini2017drone, yang2019uav}. The study on amateur drone detection \cite{uddin2020amateur} utilizes Mel-Frequency Cepstral Coefficients in the setting of supervised learning and finds that the performance degrades on low signal-to-noise ratio (SNR) settings.

Deep learning (DL) approaches significantly improved robustness: outdoor CNN-based systems outperform handcrafted features \cite{al2021audio}, and such CNN-based UAV recognition models can operate in noisy outdoor environments, but the performance degrades under strong environmental interference \cite{liu2025deep} like strong low-frequency wind and mechanical noises that overlap with drone spectral signatures. Microphone array approach \cite{jekaterynczuk2025outdoor} further revealed that microphone placement and device response introduce strong variability that impacts classification performance. DL flexibility resulted in multimodal experiments like TRIDENT \cite{Alla2025TRIDENTTR}, mixing audio models with video, radio-frequency, or infra-red video modalities and resulting in even more robust drone detection. However, existing studies rarely address cross-microphone domain shift or extreme class imbalance, both of which are common in real operational deployments.

In general audio classification, CNNs remain the common architectural choice due to their efficiency and local inductive biases. Modern CNN variants such as EfficientNet~\cite{tan2019efficientnet} and ConvNeXt~\cite{liu2022convnet} employ optimized convolutions and expanded receptive fields, achieving superior performance on spectrogram-based representations. Convolutional neural network–recurrent neural network hybrids (CRNNs) capture temporal dependencies and have been widely used for weakly labeled sound event detection problems \cite{cakir2017convolutional}. Attention-based pooling further improves clip-level detection by focusing on salient regions of the spectrogram \cite{kong2020sound}. This results in large pre-trained models like SAM-Audio \cite{Shi2025SAMAS} becoming a more common choice for zero-shot classification and segmentation task. Recent studies show that CNNs continue to outperform Transformer models on limited UAV datasets due to their stronger inductive biases and, more importantly, lower data requirements \cite{gutierrez2024comparative}. However, Transformers are also being actively introduced to the field with the AST-Drone model \cite{Zheng2025ASTDrone}.

Domain shift is a major challenge in audio machine learning. In their study on acoustic scene classification, Heittola et al. \cite{heittola2020acoustic} demonstrate significant performance drops when models are tested across microphones with different frequency responses. Recent studies incorporate explicit alignment techniques, including a two-branch network processing the audio from different sources and aligning the features from the CNN layers via specialized losses \cite{zhao2022feature}, and a CycleGAN mapping features from the source to the target domain \cite{kacprzak2021adversarial}.

PCEN provides adaptive gain control and suppresses stationary noise, thereby improving robustness under varying acoustic conditions \cite{lostanlen2018per} and for long-distance detection scenarios \cite{lostanlen2019long}. Despite its proven effectiveness, PCEN has not been systematically evaluated for UAV detection under extreme domain imbalance.

Operational UAV recordings are often scarce and imbalanced, making augmentations critical. Between-Class Learning \cite{tokozume2017learning} and Mixup \cite{zhang2017mixup} regularize decision boundaries by mixing samples, while time–frequency masking methods such as SpecAugment \cite{park2019specaugment} improve generalization by forcing models to rely on distributed acoustic cues. These methods provide useful foundations for robustness under data scarcity and motivate augmentation strategies tailored to UAV acoustics.

To sum up, prior work shows that acoustic UAV detection is feasible and that CNN-based models outperform classical methods in noisy outdoor environments. However, existing studies typically assume that training data sufficiently represent deployment conditions and report clear performance degradation under strong environmental interference, particularly low-frequency wind and mechanical noise that overlap with UAV signatures. Moreover, microphone placement and hardware response are known to affect performance substantially, yet cross-microphone generalization and extreme target-domain data scarcity in single scenarios are largely unaddressed. As a result, current literature does not fully capture the challenges of real-world UAV monitoring, where models must generalize across heterogeneous sensors and operate reliably with very limited operational data.

\section{Dataset}

The dataset consists of more than 300{,}000 audio recordings, each 9 seconds long and sampled at 32~kHz. The recordings were captured as continuous audio streams before being segmented into 9-second clips, making adjacent segments partially dependent: consecutive chunks may contain overlapping or highly similar acoustic content. The splitting procedure (Section~\ref{sec:validation}) ensures all clips from the same stream remain in a single split by construction to prevent data leakage. All samples were manually annotated using weak (file-level) labels. No timestamp-level segmentations are provided.

\begin{figure*}[t]
  \centering
  \begin{minipage}[t]{0.43\textwidth}
    \centering
    \includegraphics[width=\linewidth]{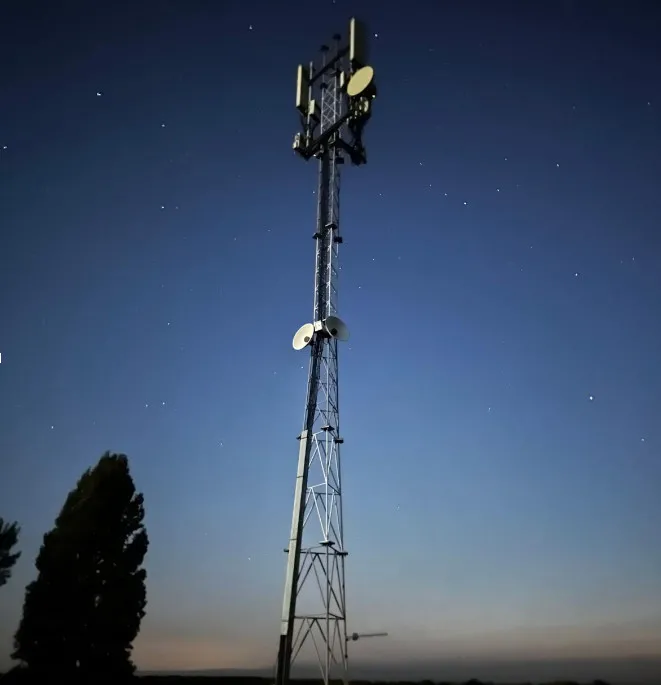}
    \captionof{figure}{Microphone-based acoustic sensing system mounted on a cell tower~\cite{speka2023acoustic}.}
    \label{fig:cell_tower_sensor}
  \end{minipage}\hspace{20px}
  \begin{minipage}[t]{0.25\textwidth}
    \centering
    \includegraphics[width=\linewidth]{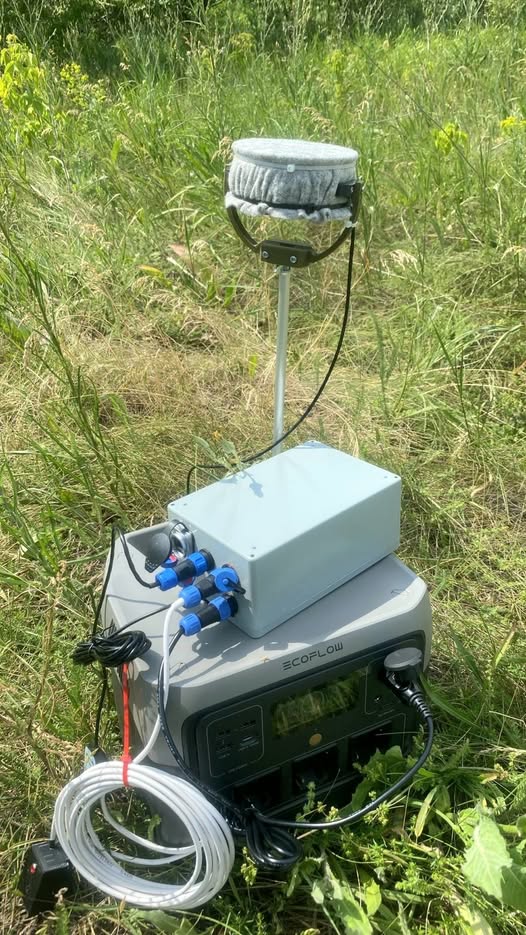}
    \captionof{figure}{Portable microphone-based acoustic sensing system.}
    \label{fig:zvook_portable_sensor}
  \end{minipage}
\end{figure*}

Data were collected using two distinct microphone types, each representing a different acoustic domain. Device Type 1 (Mic-1) is mounted on communication towers at heights ranging from 10 to 40 m (Fig. \ref{fig:cell_tower_sensor}). These recordings exhibit relatively stable environmental conditions but include characteristic noise sources such as mechanical vibrations, wind, and animal vocalizations. In addition, the device hardware introduces a distinct frequency response. Device Type 2 (Mic-2) is deployed at approximately 1 m above ground level in frontline settings (Fig. \ref{fig:zvook_portable_sensor}). This domain is more challenging, containing highly variable noise sources including human speech, music, machinery, and general battlefield clutter. As a result, Mic-2 samples exhibit broader spectral diversity and stronger nonstationary noise patterns.

% \begin{figure}[htbp]
% \centerline{\includegraphics[width=\linewidth]{images/cell_tower_sensor.png}}
% \caption{Microphone-based acoustic sensing system mounted on a cell tower~\cite{speka2023acoustic}.}
% \label{fig:cell_tower_sensor}
% \end{figure}

% \begin{figure}[htbp]
% \centerline{\includegraphics[width=\linewidth]{images/zvook_nw0.jpg}}
% \caption{Portable microphone-based acoustic sensing system.}
% \label{fig:zvook_portable_sensor}
% \end{figure}

% \begin{figure*}[t]
%   \centering
%   \begin{minipage}[t]{0.606\textwidth}
%     \centering
%     \includegraphics[width=\linewidth]{images/cell_tower_sensor.png}
%     \captionof{figure}{Microphone-based acoustic sensing system mounted on a cell tower~\cite{speka2023acoustic}.}
%     \label{fig:cell_tower_sensor}
%   \end{minipage}\hfill
%   \begin{minipage}[t]{0.354\textwidth}
%     \centering
%     \includegraphics[width=\linewidth]{images/zvook_nw0.jpg}
%     \captionof{figure}{Portable microphone-based acoustic sensing system.}
%     \label{fig:zvook_portable_sensor}
%   \end{minipage}
% \end{figure*}

The Mic-1 dataset contains four sound classes of interest in addition to a background environmental noise class:
\begin{itemize}
    \item \textbf{Drones} – predominantly Shahed-type and other long-range UAVs powered by electric or internal combustion engines.
    \item \textbf{Jet Aircraft} – aircraft powered by jet turbines.
    \item \textbf{Helicopters} – rotary-wing aircraft.
    \item \textbf{Propeller Aircraft} – light aircraft with piston or turboprop engines driven by propellers.
    \item \textbf{Environmental Noise} – background recordings including wind, animals, anthropogenic sounds, and other non-target noise sources.
\end{itemize}

The Mic-2 dataset contains a single sound class of interest along with a background environmental noise class:
\begin{itemize}
    \item \textbf{Small Drone} – predominantly radio-guided first-person view (FPV) drones and DJI Mavic platforms, including night bombers, fiber-optic FPVs, and other small-to-medium UAVs operating at short to medium ranges. This class aggregates multiple drone types, as it is often infeasible to reliably distinguish between them based solely on human perception.
    \item \textbf{Environmental Noise} – same as \textbf{Environmental Noise} for Mic-1.
\end{itemize}

It is important to note that the \textbf{Environmental Noise} class differs substantially between Mic-1 and Mic-2. This domain-specific discrepancy is analyzed in detail in Section \ref{sec:domain_shift}.

Mic-2 represents the target domain, while Mic-1 serves as an auxiliary domain. The training dataset may contain recordings from only Mic-2 or both devices, whereas the validation and test sets are evaluated on both Mic-1 and Mic-2 data and their corresponding classes. As shown in Table~\ref{table:class-distr}, two datasets are considered in this study: a \emph{small} set and a \emph{full} set. The small set is a subset of the full set and contains approximately 10.5$\times$ fewer \textbf{Small drone} samples, 2.1$\times$ fewer Mic-2 \textbf{Noise} samples and 6$\times$ fewer Mic-1 samples. This small set represents a typical early-stage development scenario for acoustic detection algorithms, in which only a limited amount of data is available for model training; therefore, it is used as the primary dataset in this research. The best-performing model is additionally retrained on the full set to evaluate performance in a setting where a larger training dataset is available. 

For the small set, the label distribution across devices is highly imbalanced: Mic-1 accounts for more than 99\% of the training data, while Mic-2 contains approximately 3{,}000 samples ($<1\%$). Moreover, only 441 samples correspond to the \textbf{Small drone} class, making it extremely rare. This class does not appear in Mic-1, resulting in a partially disjoint label space. A complete per-class, per-device distribution is presented in Table~\ref{table:class-distr}.

\begin{figure}[t]
  \centering
  \includegraphics[width=0.9\linewidth]{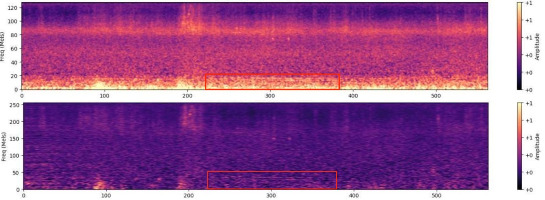}
  \caption{Example of a Mel spectrogram (top) and PCEN representation (bottom) for a UAV audio sample.}
  \label{fig:spec_and_pcen}
\end{figure}

\begin{table}[htbp]
\caption{Train classes distribution}
\begin{center}
\begin{tabular}{|c|c|c|c|c|}
\hline
\multirow{2}{*}{\textbf{Train set}} & \multirow{2}{*}{\textbf{Device}} & \multirow{2}{*}{\textbf{Class}} & \multicolumn{2}{c|}{\textbf{Samples statistics}} \\
\cline{4-5} 
& & & \textbf{\textit{Count}} & \textbf{\textit{Length (hours)}} \\
\hline
\multirow{7}{*}{Small} & \multirow{5}{*}{Mic-1} & Drone & 4758 & 11.9 \\
 & & Jet Aircraft & 6012 & 15.0 \\
 & & Helicopter & 2559 & 6.4 \\
 & & Propeller Aircraft & 1797 & 4.5 \\
 & & Noise & 50000 & 125.0 \\
 \cline{2-5}
 & \multirow{2}{*}{Mic-2} & Small Drone & 441 & 1.1 \\
 & & Noise & 2519 & 6.3 \\
\hline
\multirow{7}{*}{Full} & \multirow{5}{*}{Mic-1} & Drone & 5421 & 13.7 \\
 & & Jet Aircraft & 7141 & 18.1 \\
 & & Helicopter & 3225 & 8.0 \\
 & & Propeller Aircraft & 2194 & 5.6 \\
 & & Noise & 304919 & 761.2 \\ 
 \cline{2-5}
 & \multirow{2}{*}{Mic-2} & Small Drone & 4648 & 11.6 \\
 & & Noise & 5316 & 13.1 \\
\hline
\end{tabular}
\end{center}
\label{table:class-distr}
\end{table}

Representative Mel and PCEN spectrograms for UAV signals are shown in Figure~\ref{fig:spec_and_pcen}.

\section{Validation protocol}
\label{sec:validation}

We opt out of $K$-fold cross-validation to reduce computational cost and instead evaluate models using fixed validation and test splits based on time and location. The three sets are completely time independent.

\begin{itemize}
    \item \textbf{Training Set:} Collected during controlled field tests in non-combat zones. This represents the initial data available during development.
    \item \textbf{Validation Set:} Collected during the \textbf{Winter season}. This set contains recordings from both active combat zones and field tests. We use this set to tune hyperparameters and check for overfitting.
    \item \textbf{Test Set:} Collected during the \textbf{Spring season}, strictly from active combat zones. This is the final evaluation set to see how the model performs in a new season and in unseen locations.
\end{itemize}

Validation and test set class distributions are depicted in Table~\ref{table:val-test-class-distr}.

Although the primary objective is binary detection (\textit{Small Drone} vs. All other classes), we select the best model checkpoint using the Macro F1-score calculated across all classes, including the auxiliary categories from Mic-1. However, we also track the binary metric, ensuring the model does not overfit to auxiliary classes.

This strategy addresses the extreme data scarcity of the target class in the validation set. Relying exclusively on binary metrics for such a small sample size can lead to high variance in model selection. By optimizing for performance across a broader range of acoustic sources, we ensure the backbone learns robust, discriminative spectral features. This serves as a proxy for generalization capability, leading to more reliable detection of the target class.

For the final results on the Test and Validation sets, we report Binary Metrics (Small Drone vs. All other classes): Precision, Recall, and F1-score.

\begin{table}[htbp]
\caption{Validation and test classes distribution}
\begin{center}
\begin{tabular}{|c|c|c|c|c|}
\hline
\multirow{2}{*}{\textbf{Set}} & \multirow{2}{*}{\textbf{Device}} & \multirow{2}{*}{\textbf{Class}} & \multicolumn{2}{c|}{\textbf{Samples statistics}} \\
\cline{4-5} 
& & & \textbf{\textit{Count}} & \textbf{\textit{Length (hours)}} \\
\hline
\multirow{7}{*}{Val} & \multirow{5}{*}{Mic-1} & Drone & 1969 & 4.92 \\
 & & Jet Aircraft & 1502 & 3.75 \\
 & & Helicopter & 2800 & 7.0 \\
 & & Propeller Aircraft & 412 & 1.03 \\
 & & Noise & 10765 & 26.91 \\
 \cline{2-5}
 & \multirow{2}{*}{Mic-2} & Small Drone & 805 & 2.01 \\
 & & Noise & 343 & 0.86 \\
\hline
% \multirow{7}{*}{Test} & \multirow{5}{*}{Mic-1} & Drone & 4758 & 11.9 \\
%  & & Jet Aircraft & 6012 & 15.0 \\
%  & & Helicopter & 2559 & 6.4 \\
%  & & Propeller Aircraft & 1797 & 4.5 \\
%  & & Noise & 50000 & 125.0 \\
% \cline{2-5}
\multirow{7}{*}{Test} & \multirow{5}{*}{Mic-1} & Drone & 32876 & 82.19 \\
 & & Jet Aircraft & 2217 & 5.54 \\
 & & Helicopter & 2408 & 6.02 \\
 & & Propeller Aircraft & 662 & 1.66 \\
 & & Noise & 11496 & 28.73 \\
 \cline{2-5}
 & \multirow{2}{*}{Mic-2} & Small Drone & 2545 & 6.13 \\
 & & Noise & 2549 & 6.28 \\
\hline
\end{tabular}
\end{center}
\label{table:val-test-class-distr}
\end{table}

\section{Domain Shift}
\label{sec:domain_shift}

\begin{figure*}[t]
  \centering
  % -------- First row --------
  \begin{minipage}[t]{0.49\textwidth}
    \centering
    \includegraphics[width=\linewidth]{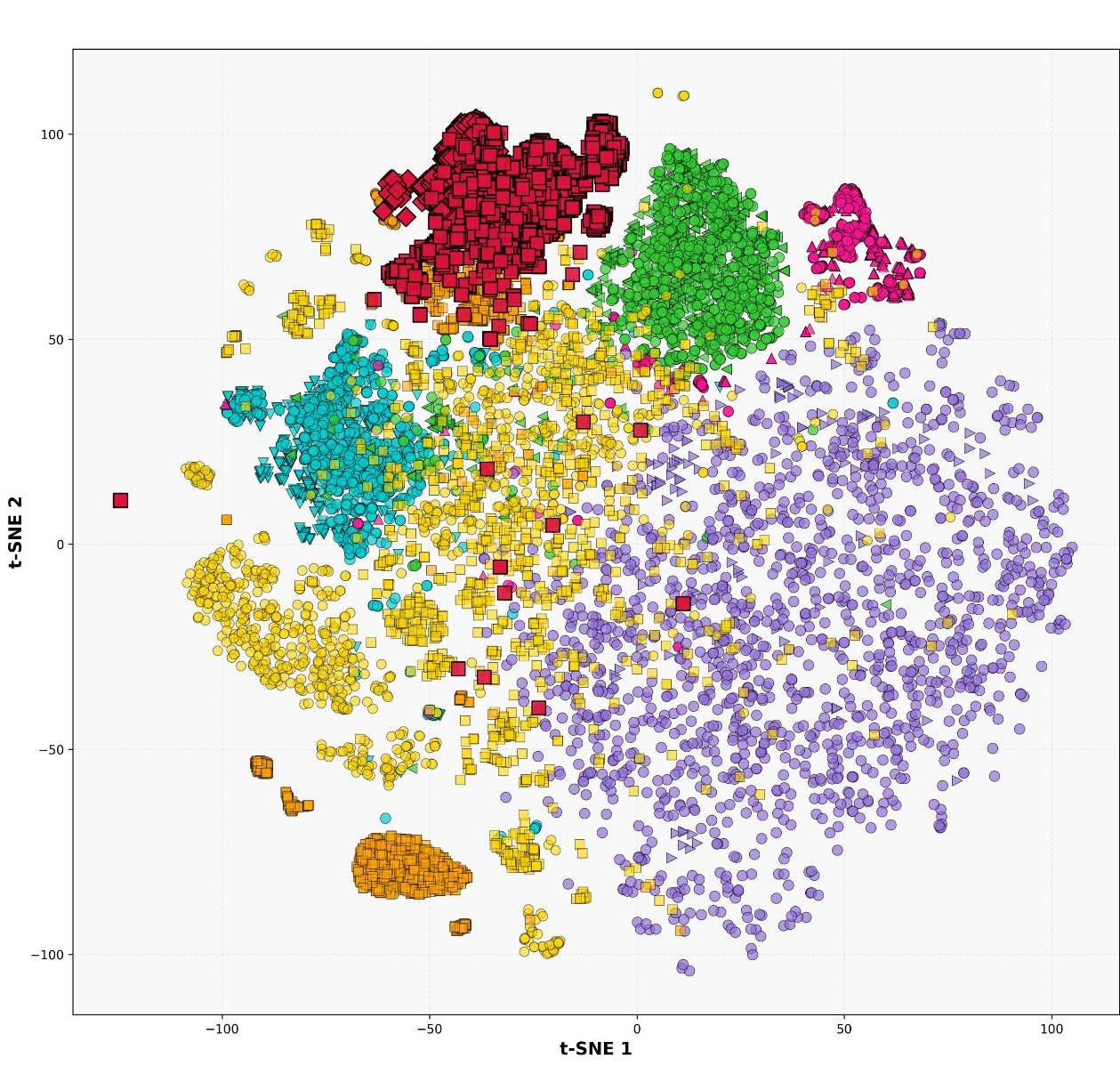}
    \captionof{figure}{t-SNE projection of model embeddings obtained from the enhanced model (Log-Mel).}
    \label{fig:tsne_best}
  \end{minipage}\hfill
  \begin{minipage}[t]{0.49\textwidth}
    \centering
    \includegraphics[width=\linewidth]{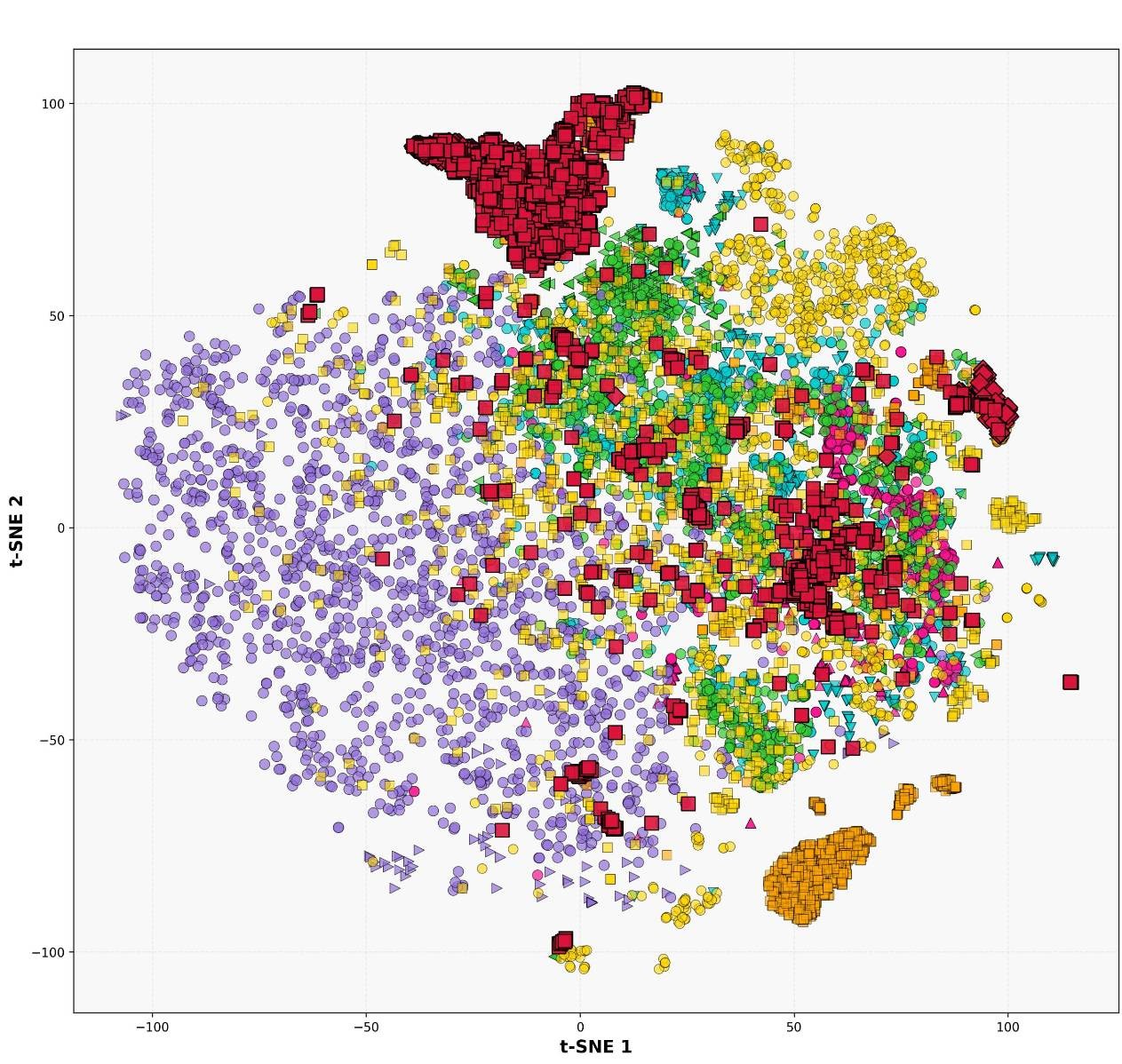}
    \captionof{figure}{t-SNE projection of model embeddings obtained from the baseline model.}
    \label{fig:tsne_baseline}
  \end{minipage}
  
  % --- Shared Legend ---
  \centering
  \small
  \textbf{Common legend} (Validation / Test set signs): \quad
  % Classes
  \textcolor[HTML]{DB2C4D}{$\blacklozenge/\blacksquare$} Small Drone \quad 
  \textcolor[HTML]{2ED8D5}{$\blacktriangledown/\bullet$} Jet \quad 
  \textcolor[HTML]{FF5CB8}{$\blacktriangle/\bullet$} Propeller Aircraft \quad
  \textcolor[HTML]{73D372}{$\blacktriangleleft/\bullet$} Helicopter \quad
  \textcolor[HTML]{AB91EA}{$\blacktriangleright/\bullet$} Drone \quad
  \textcolor[HTML]{ECDD4E}{$\bullet/\blacksquare$} Environmental Noise (Mic 1) \quad 
  \textcolor[HTML]{FFD25B}{$\bullet/\blacksquare$} Environmental Noise (Mic 2)

  \vspace{0.5em}

  % -------- Second row --------
  \begin{minipage}[t]{0.49\textwidth}
    \centering
    \includegraphics[width=\linewidth]{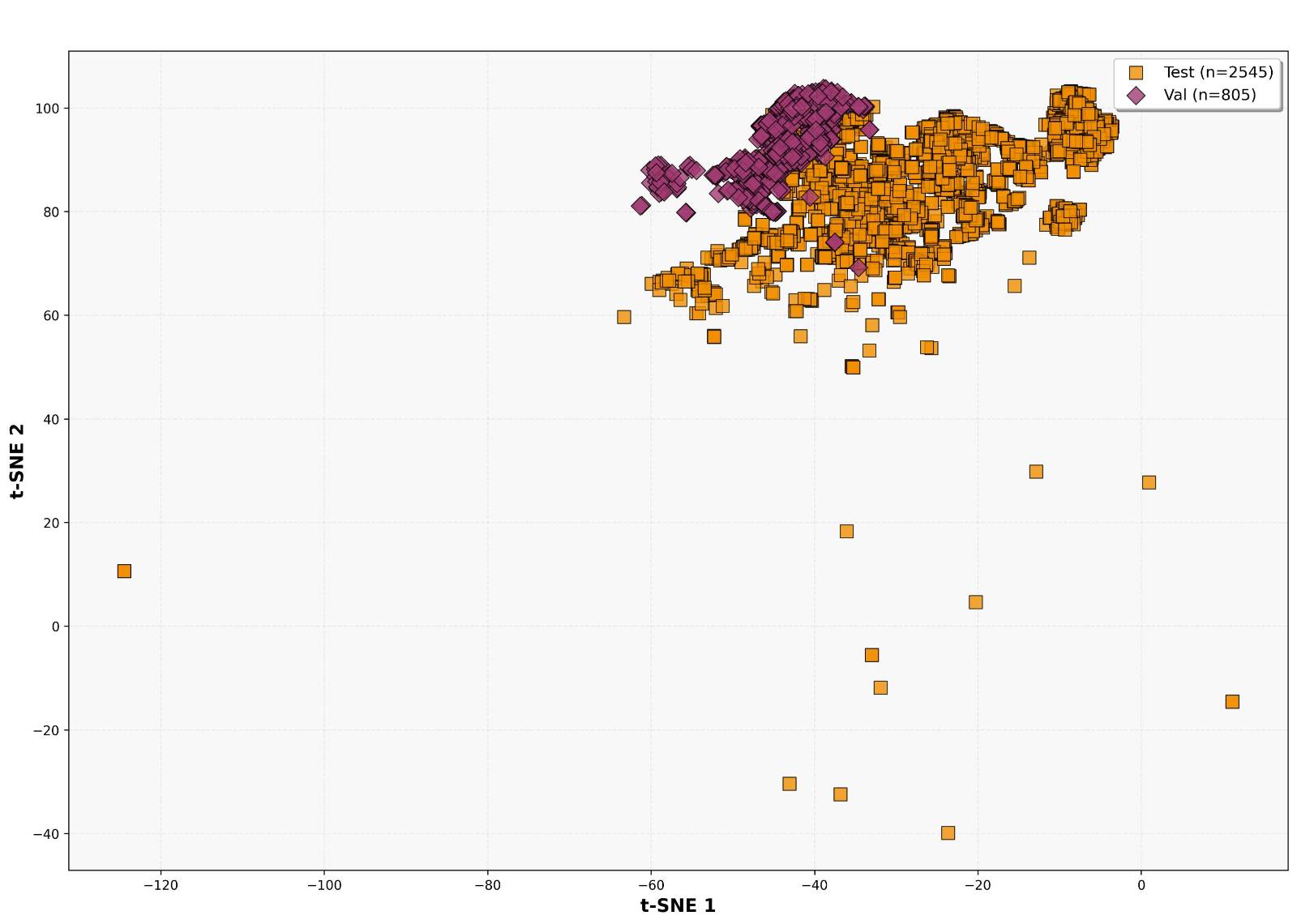}
    \captionof{figure}{t-SNE projection of model embeddings of class "Small Drone" obtained from the enhanced model (Log-Mel) for the validation and test sets.}
    \label{fig:tsne_val_test}
  \end{minipage}\hfill
  \begin{minipage}[t]{0.49\textwidth}
    \centering
    \includegraphics[width=\linewidth]{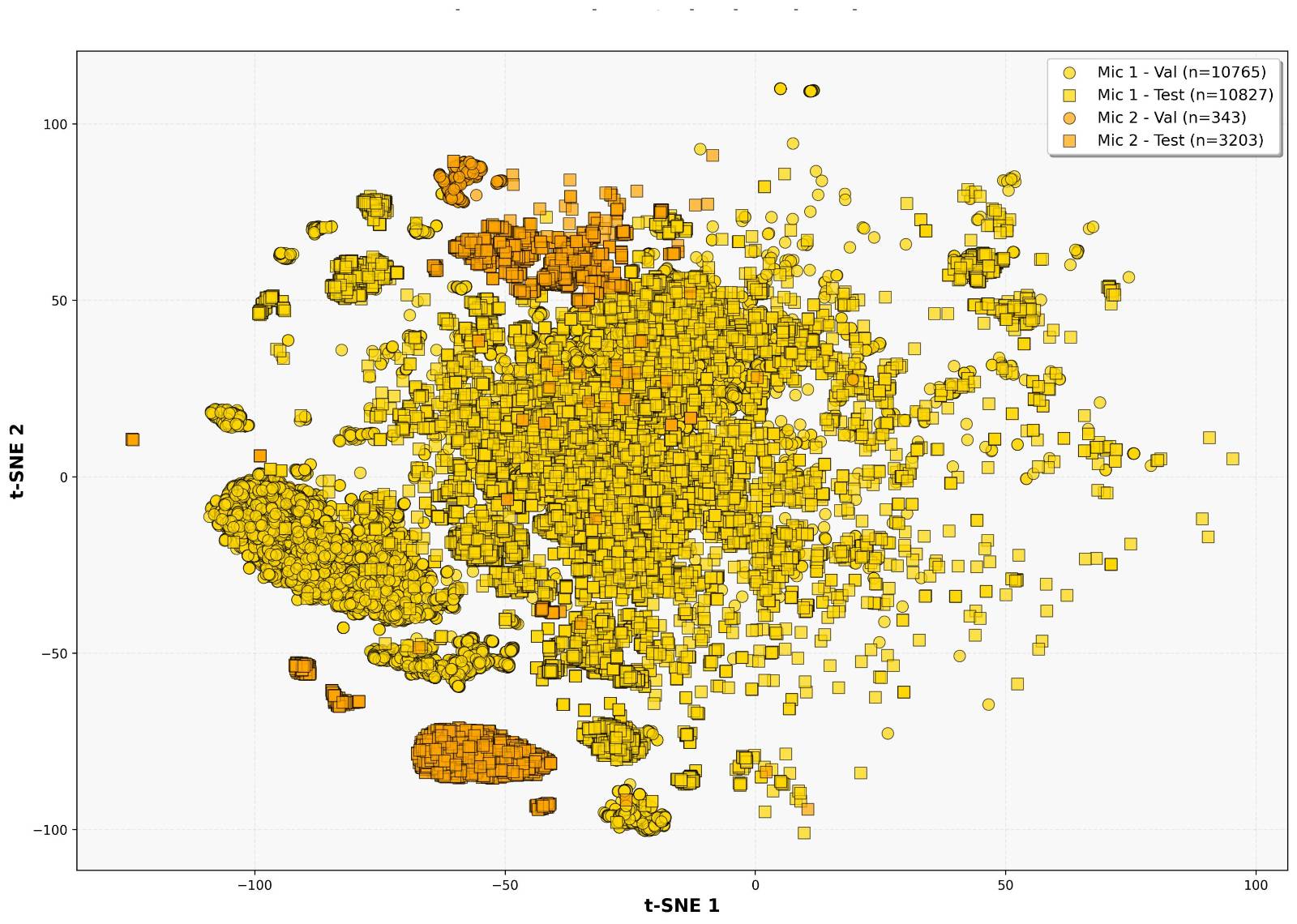}
    \captionof{figure}{t-SNE projection of model embeddings of class "Noise" obtained from the enhanced model (Log-Mel) for the validation and test sets and for Mic-1 and Mic-2.}
    \label{fig:tsne_mics}
  \end{minipage}
\end{figure*}

To illustrate domain differences in a cross-device setting, we provide t-SNE~\cite{Linderman2017FastIT} projections of model embeddings for a data subset. The embedding spaces for the baseline model and for the model with all proposed enhancements are shown in Figures~\ref{fig:tsne_best} and~\ref{fig:tsne_baseline}. Embeddings from all classes form distinct clusters, indicating that the enhanced model reasonably separates the data space; however, \textbf{Environmental Noise} from Mic-2 forms a distinct cluster, which clearly illustrates the presence of domain shift between the two domains. While noises from the two domains are partially separated from each other (Fig.~\ref{fig:tsne_mics}), they remain well separated from other classes. This suggests that although the domain shift problem is not fully resolved, it should not significantly affect model performance in this particular scenario. In contrast, the same projections for the baseline model show that the embedding space is much less separable. Additionally, embeddings from the validation and test sets are also separable and form two distinct clusters (Fig.~\ref{fig:tsne_val_test}).

\begin{table}[htbp]
\caption{Analysis of domain invariance with respect to seasonal and sensor variability (lower clustering metrics correspond to higher invariance)}
\begin{center}
\begin{tabular}{|l|l|c|c|}
\hline
\textbf{Shift Type} & \textbf{Model Frontend} & \textbf{Sil. ($\downarrow$)} & \textbf{$k$-NN (\%) ($\downarrow$)} \\
\hline
\multirow{3}{*}{\shortstack[l]{Global Noise \\ Sensor Shift \\ (Mic-1 vs. Mic-2)}} & Baseline (Log-Mel) & \textbf{0.1346} & 97.56 \\
& Enhanced (Log-Mel) & 0.2084 & 97.84 \\
& Enhanced (PCEN) & 0.1700 & \textbf{96.13} \\
\hline
\multirow{3}{*}{\shortstack[l]{Seasonal Shift \\ (Winter vs. Spring)}} & Baseline (Log-Mel) & 0.0093 & 74.60 \\
& Enhanced (Log-Mel) & 0.0086 & 72.94 \\
& Enhanced (PCEN) & \textbf{0.0048} & \textbf{70.85} \\
\hline
\multirow{3}{*}{\shortstack[l]{Mic-2 Noise \\ Seasonal Shift \\ (Winter vs. Spring)}} & Baseline (Log-Mel) & 0.3851 & 98.79 \\
& Enhanced (Log-Mel) & 0.3072 & 95.96 \\
& Enhanced (PCEN) & \textbf{0.2592} & \textbf{93.91} \\
\hline
\end{tabular}
\label{table:domain_invariance}
\end{center}
\end{table}

Table~\ref{table:domain_invariance} evaluates the framework's robustness across sensor and seasonal shifts via quantitative metrics such as the Silhouette Coefficient (Sil.) \cite{rousseeuw1987silhouettes} and \textit{k}-NN cluster purity (\textit{k}=10, cosine similarity).
The Global Noise Sensor Shift experiment unites validation and test noise samples, forming two groups based on the microphone difference. The metrics suggest that the enhanced models struggle to align the domains, as the clusters appear more distinct compared to the baseline model. 

The results for Seasonal Shift, calculated by comparing validation and test distributions, hint at the absence of a seasonal shift for the signals. 

In contrast, a significant shift is present between noise classes. Mic-2 Noise Seasonal Shift metrics indicate that the baseline model distinguishes between environmental noise classes with high clarity (Sil. 0.3851, \textit{k}-NN $> 98\%$). The Enhanced (PCEN) model mitigates this effect by design via blurring the non-stationary environmental textures, reducing the silhouette to 0.2592 and \textit{k}-NN purity to 93.91\%, which suggests a higher cluster overlap.

We additionally perform adversarial validation \cite{Pan2020AdversarialVA} by training two model variants to distinguish noise samples originating from two different sources. The corresponding validation results are reported in Table~\ref{table:adv-results}. We consider two setups: a baseline modeling approach described in Section~\ref{sec:Method}, and an enhanced variant used throughout this work. The baseline model achieves ROC AUC close to 100\% on both the validation and test sets, indicating near-perfect separability of samples recorded with different microphones. In contrast, the enhanced model largely loses this discriminative ability, yielding an ROC AUC of approximately 70\% on the test set. While this performance remains above random guessing, it suggests that a substantial portion of the audio samples have become more similar with respect to microphone-specific characteristics. While the validation ROC AUC for the enhanced model remains close to 100\%, we hypothesize that this is due to the limited number of noise classes available for Mic-2, which results in reduced data diversity.

\begin{table}[htbp]
\caption{Adversarial validation results}
\begin{center}
\begin{tabularx}{\columnwidth}{|X|c|c|c|c|}
\hline
\multirow{2}{*}{\textbf{Model setup}} & \multicolumn{2}{c|}{\textbf{Validation set}} & \multicolumn{2}{c|}{\textbf{Test set}} \\
\cline{2-5} 
& \textbf{\textit{ROC AUC}} & \textbf{\textit{F1}} & \textbf{\textit{ROC AUC}} & \textbf{\textit{F1}} \\
\hline
Baseline model setup & 99.99\% & 100\% & 97.07\% & 66\% \\
\hline
Enhanced model setup based on Section~\ref{sec:Method} \ & 99.67\% & 97\% & 69.1\% & 62\% \\
\hline
\end{tabularx}
\end{center}
\label{table:adv-results}
\end{table}

\section{Method}
\label{sec:Method}
Our framework follows the architecture proposed in BirdCLEF+ 2025 second-place solution \cite{sydorskyi2025tackling}. Similar to the proposed approach, we employ a ConvNeXt-Tiny~\cite{liu2022convnet} backbone as a feature extractor and an sound event detection (SED)-style attention head to accommodate weak labels. 

In low-SNR scenarios, the standard pooling strategy can lead to destructive aggregation: high-energy low-frequency noise may overlap with the UAV signal, effectively masking discriminative features. To address this issue, we remove the Generalized Mean (GeM) frequency pooling layer and replace it with a learnable frequency projector implemented as a convolutional block, with convolutions applied along the frequency dimension. Unlike pooling, which collapses spectral resolution, this projection preserves convolved frequency structure and enables the model to learn non-linear, weighted combinations of frequency bands.

For multiclass classification, we remove the \texttt{tanh} constraint from the attention mechanism, allowing unbounded scores that sharpen temporal focus. We also replace the final sigmoid with a class-wise softmax on the aggregated output to enforce mutual exclusivity between classes.

\subsection{Audio Representation}

All inputs are converted into PCEN \cite{lostanlen2018per} spectrograms. We experimented with PCEN to address the extreme dynamic range variance caused by environmental factors. PCEN suppresses the stationary background noise and normalizes loudness variations, which improves the visibility of weak spectral structures that are otherwise dominated by persistent noise patterns, making the representation especially robust to high-energy low-frequency noise. 

\subsection{Data Augmentation}

To address the severe data imbalance, particularly in Mic-2, we apply a broad set of augmentations. The following transformations operate directly on the waveform:
\begin{itemize}
    \item Time stretching.
    \item Random gain.
    \item Waveform reversal.
    \item Spec augment-style masking \cite{park2019specaugment}.
\end{itemize}
These augmentations introduce temporal and amplitude variability without drastically altering the underlying spectral signatures.

We further employ noise injection via mixup \cite{zhang2017mixup}, where noise samples are mixed into training audio while preserving the original class label. This serves two purposes: (1) generating additional training examples for the minority domain, and (2) gradually pulling the Mic-1 representation toward the Mic-2 acoustic space.

A second critical augmentation is the random removal of the segment with the highest root mean square (RMS) amplitude for all classes except noise. We locate a 2-second region of maximum energy and randomly replace 0.5–1.0 seconds inside it with noise, following a curriculum that starts from 0.3–0.5 s in early epochs and increases to 0.5–1.0s . This forces the model to focus on other features besides the loudest chunk, since the signal is almost guaranteed to exist around it.

\subsection{Training Setup}

The model is trained on the combined dataset from both microphones. Classes from  Mic-1 are used as auxiliary classes, enabling the backbone and attention head to learn generalizable acoustic structure while still optimizing the model for the primary task of distinguishing \textbf{Small drone} vs. all other classes.

\section{Experiments and Discussion}

\begin{table*}[htbp]
\caption{Experiment results (best metric in category being \textbf{semibold} and the following one \underline{underlined})}
\begin{center}
\begin{tabularx}{0.9\textwidth}{|X|c|c|c|c|c|c|c|}
\hline
\multirow{2}{*}{\textbf{Model}} &
\textbf{Parameters} &
\multicolumn{3}{c|}{\textbf{Small drone metrics on validation}} &
\multicolumn{3}{c|}{\textbf{Small drone metrics on test set}} \\
\cline{3-8}
& \textbf{(M)} &
\textbf{\textit{F1, \%}} &
\textbf{\textit{Precision, \%}} &
\textbf{\textit{Recall, \%}} &
\textbf{\textit{F1, \%}} &
\textbf{\textit{Precision, \%}} &
\textbf{\textit{Recall, \%}} \\
\hline

Zvook \cite{zvookwebsite} Baseline & 37 &
96.35 & 99.21 & 93.66 &
69.10 & 81.59 & 59.92 \\
\hline

Baseline on Mic2 & 28.22 &
17.00 & 88.00 & 0.094 &
17.40 & 87.70 & 0.097 \\

Baseline on Mic2 + Mic1 & 28.22 &
91.40 & \textbf{100.00} & 84.10 &
49.80 & 88.80 & 34.60 \\

\textit{+ SWA} & 28.22 &
96.20 & \textbf{100.00} & 92.70 &
59.70 & 87.20 & 45.30 \\

~\textit{+ Mixup} & 28.22 &
83.20 & 72.60 & 97.40 &
75.60 & 74.40 & \textbf{76.90} \\

~~\textit{+ RMS based masking} & 28.22 &
90.60 & 84.40 & \textbf{97.80} &
75.20 & 78.10 & 72.50 \\

~~~\textit{+ Projector} & 30.97 &
96.10 & 95.80 & 96.40 &
77.80 & \textbf{92.70} & 67.10 \\

~~~~\textit{+ Using full dataset} & 30.97 &
\textbf{97.70} & \underline{98.10} & \underline{97.30} &
77.10 & 90.90 & 67.00 \\

~~~\textit{+ PCEN} & 28.22 &
\underline{97.50} & 97.90 & 97.10 &
74.60 & 87.30 & 65.10 \\

~~~~\textit{+ Projector} & 30.97 &
96.10 & 96.80 & 95.30 &
\underline{77.90} & \underline{92.20} & 67.50 \\

~~~~~\textit{+ Using full dataset} & 30.97 &
97.10 & 98.00 & 96.10 & 
\textbf{78.60} & 84.30 & \underline{73.60} \\

\hline

Ensemble (Small training set) & 61.94 &
97.23 & 98.60 & 95.90 &
81.14 & \textbf{96.18} & 70.18 \\

Ensemble (Full training set) & 61.94 &
\textbf{98.18} & \textbf{99.36} & \textbf{97.02} &
\textbf{82.22} & 92.70 & \textbf{73.87} \\

\hline

SAM-Audio-small \cite{Shi2025SAMAS} & 5{,}807 &
\underline{10.60} & \underline{5.72} & \textbf{72.67} &
4.45 & 2.38 & \underline{34.05} \\

AST-Drone \cite{Zheng2025ASTDrone} & 86 &
\textbf{42.82} & \textbf{32.40} & \underline{63.11} &
\textbf{55.36} & \textbf{49.02} & \textbf{63.58} \\

TRIDENT \cite{Alla2025TRIDENTTR}  LeNet-Audio \cite{Szegedy2014GoingDW} & 0.012 &
1.68 & 2.20 & 1.37 & \underline{19.29} & \underline{27.84} & 14.75 \\
\hline
\end{tabularx}
\end{center}
\label{table:exp-results}
\end{table*}

\subsection{External Baseline Models}

Several open-source and open-weight models were selected and taken as-is as baselines for our experiments:
\begin{itemize}
    \item \textbf{SAM Audio} \cite{Shi2025SAMAS} is a recent foundational model for audio segmentation. It is included to assess the feasibility of zero-shot UAV detection. The model was evaluated in FP16 inference mode using the prompt \texttt{Small UAV FPV flight}. The resulting target-only audio was post-processed by thresholding its maximum absolute amplitude to obtain a clip-level class decision based on detected segments. This approach was selected among several alternatives that relied on combinations of extracted target and residual signals. However, its main objective is unrelated to military domain, which likely led to the relatively poor performance (Table~\ref{table:exp-results}).
    \item \textbf{AST-Drone} \cite{Zheng2025ASTDrone} is a fine-tuned variant of the Audio Spectrogram Transformer (AST) \cite{Gong2021ASTAS}, specifically adapted for UAV detection. It represents one of the largest publicly available models tailored to this task. It performed the best among open source baselines due to its capacity and clever architecture, still lacking real-world training data and cross-environment generalization.
    \item \textbf{TRIDENT LeNet-Audio} \cite{Alla2025TRIDENTTR, Szegedy2014GoingDW} is a lightweight audio model originating from a multimodal UAV detection framework. It is included to evaluate the standalone performance of a compact, resource-efficient UAV detection model and to compare it against larger and more specialized approaches. Unfortunately, its compact size led to inability to handle drone detection scenarios effectively (Table~\ref{table:exp-results}).
\end{itemize}

Additionally, a proprietary 5-fold ensemble model from Zvook \cite{zvookwebsite} trained on the same full training set was used as a main baseline for the paper.

\subsection{Detailed Training Setup}

The extreme class imbalance (with the target class comprising $<1\%$ of the data) poses a risk of the model collapsing into a trivial noise detector. To mitigate this, we employed a square-root sampling strategy with replacement in the data loader. Crucially, while environmental noise from both domains is mapped to a single class label, we sample them independently based on their respective counts.

We implemented a diverse augmentation pipeline designed to simulate the harsh acoustic conditions of the target domain. We applied a range of augmentations to simulate physical variability, including random time shifting ($\pm 100$ms), sinusoidal time warping, random gain ($\pm 3$dB), and waveform reversal ($p=0.5$). Unlike standard Mixup, we employed a noise-injection strategy where training samples are mixed exclusively with noise from either the source (Mic-1) or target (Mic-2) domain with equal probability. The mixing coefficient $\lambda$ is capped at $0.3$, and the original label is retained. This effectively synthesizes ``low-SNR'' examples, forcing the model to recognize drone signatures embedded in target-domain noise. Qualitative analysis revealed that baseline models often over-relied on the highest-energy segment of a clip. To counter this, we locate the 2-second region of maximum energy and randomly mask a sub-segment ($0.3$s--$1.0$s) within it via a curriculum starting from epoch 5 linearly increasing the interval. This prevents early metric collapse while forcing the model to attend to weaker, redundant harmonic features once stable features have been learned.

All models were implemented using PyTorch Lightning. We utilized a ConvNeXt-Tiny~\cite{liu2022convnet} backbone pretrained on ImageNet-1K. Audio inputs ($32$~kHz, $9$ seconds) were converted to spectrograms using a hop length of $512$ samples. For the Log-Mel Spectrogram baselines, we utilized a window size of $2048$ samples and $128$ Mel bands. For the PCEN/Log-Mel configuration, we employ a higher-resolution setup with a window size of $4096$ samples and $256$ bands. We utilized a PCEN frontend with fixed initialization parameters: smoothing $s=0.015$, gain $\alpha=0.8$, bias $\delta=2.0$, power $r=0.5$, and compression power of $2.0$. Training was conducted for $100$ epochs with a batch size of $64$ using the AdamW optimizer ($lr=1\text{e-}4$, $\beta=(0.9, 0.999)$, $\epsilon=1\text{e-}8$). We employed a Cosine Annealing scheduler that decayed the learning rate to $1\text{e-}6$ over the full training duration. Focal loss with parameters $\alpha = 1$ and $\gamma = 2$ is used as a loss function.

\subsection{Experiments}

We establish the baseline using the setup above without domain adaptation techniques. Training solely on the available target-domain data (Mic-2) resulted in model collapse, demonstrating that the limited data is insufficient for the model to generalize.

The quantitative results of our incremental ablation study are presented in Table~\ref{table:exp-results}. The inclusion of auxiliary aircraft classes serves two critical functions. These classes occupy similar frequency bands to small drones. By forcing the model to discriminate between overlapping harmonic structures, we prevent it from learning simple energy-based features. The auxiliary classes are also better represented in the training set, which provides a richer feature space during training.

However, this strategy inherently assumes a beneficial spectral overlap. If this assumption fails, the model risks negative transfer, which includes wasting representational capacity on unrelated features, diluting target confidence if auxiliary classes are too similar to the drone, and overfitting to domain-specific noise. We explicitly mitigate these risks through the representation and augmentation strategies detailed below.

The introduction of noise-injection Mixup aligned the noise distributions of the two domains, preventing the model from memorizing the specific background texture of the source domain, while exposing the model to target domain scenarios. Curriculum RMS masking is utilized to address the model's tendency to over-rely on the highest-energy segment by forcing attention to weaker harmonic features distributed throughout the clip.

In low-SNR scenarios with strong low-frequency interference, the fundamental drone frequency may be masked. Standard Global Average Pooling collapses all frequency information into a single scalar, destroying the model's ability to distinguish between the frequency bands for detection. The frequency projector applies learned 1D convolutions along the frequency axis, enabling the model to detect harmonic ratios even when the fundamental frequency is occluded, learn frequency-band-specific suppression of stationary noise, and preserve spectral topology for non-linear feature extraction. This is particularly critical for Mic-2 deployment, where ground-level recording introduces stronger low-frequency environmental interference compared to tower-mounted Mic-1 sensors.

Given the aggressive augmentation and sampling strategies, we observed high variance in validation metrics between epochs. Stochastic Weight Averaging \cite{izmailov2018averaging} is applied to smooth that effect. We utilized a top-5 checkpoint selection strategy based on Macro-F1 to ensure the final model was robust.

PCEN acted effectively as a background subtractor. In high-noise samples, PCEN suppressed stationary wind rumble, revealing spectral lines invisible in the Log-Mel spectrogram (Fig.~\ref{fig:spec_and_pcen}). The distinct failure modes of PCEN (sensitive to tuning) and Mel-spectrograms (sensitive to gain) proved complementary. An ensemble of the two modalities on the full dataset yielded 82.22\% F1. The model trained on small dataset achieved 81.14\% F1, outperforming Zvook baseline model by a large margin, while utilizing 5x less target domain samples. Overall, we have achieved an F1 score increase from $55.4\%$ (best score of the open-source AST-Drone model) up to $82.2\%$ (score of our ensemble model on the full training set).

\subsection{Detection Range Analysis}

The original dataset contains a subdivision of the Small drone class based on perceived distance to the sensor, as annotated by human labelers. Although this perception can be influenced by physical obstacles, microphone characteristics, and drone type, it provides the only practical approximation of sensor-to-drone distance available in a real combat scenario.

\begin{figure}[t]
  \centering
  \includegraphics[width=0.6\linewidth]{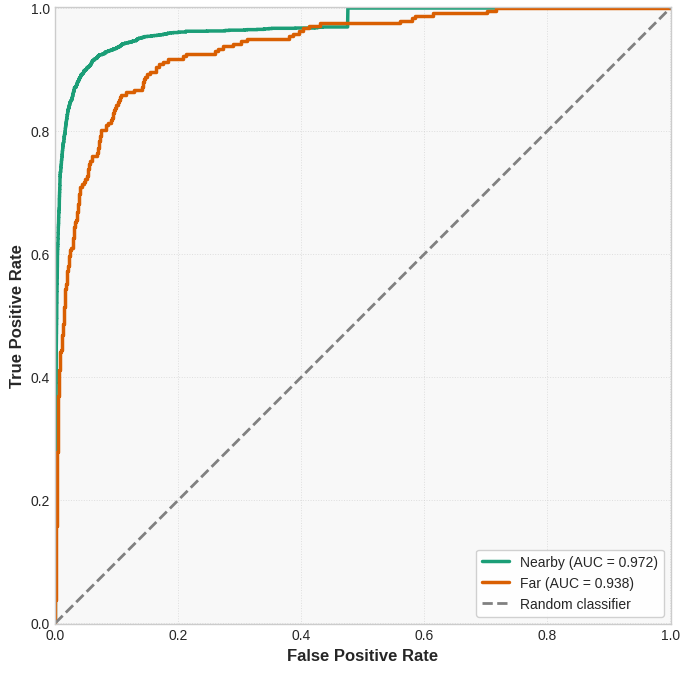}
  \caption{ROC curves for different drone distance subgroups and the final ensemble model.}
  \label{fig:roc_curves_dists}
\end{figure}

This subdivision was used to evaluate the final ensemble model performance across different distance ranges using ROC curves (Fig.~\ref{fig:roc_curves_dists}) and the ROC AUC metric. As a result, $2304$ recordings labeled as nearby or medium-distance flybys achieve an ROC AUC of $97.2\%$, while $241$ recordings labeled as far-distance flybys achieve an ROC AUC of $93.8\%$. 

The performance drop for distant flybys is expected due to lower signal-to-noise ratios and weaker acoustic signatures at larger distances. Nevertheless, the model maintains high detection accuracy across both subgroups, indicating robust performance under varying detection ranges.

\section{Conclusion}

Acoustic UAV detection across heterogeneous recording devices remains largely underexplored. We proposed a domain-aware acoustic classification framework combining PCEN representations, attention-based pooling, and noise-driven curriculum augmentations. Evaluations against multiple open-source and proprietary baselines have shown consistent improvements in cross-device generalization and detection accuracy.

Despite these advancements, several avenues for refinement remain:
\begin{itemize}
    \item Current models treat UAVs as a monolithic class. Developing a fine-grained taxonomy (e.g., distinguishing between multi-rotors like DJI Mavics and high-speed FPV drones) is essential for facing tactical problems.
    \item Our approach processes audio in independent segments. Integrating information from sequential predictions could improve detection stability and reduce false-positive rates.
    \item The lack of explicit spatial data (distance and altitude) limits our ability to perform detailed error analysis and performance reporting.
\end{itemize}

Consequently, future research should focus on:
\begin{itemize}
    \item Incorporating additional metadata into the dataset, including distance metrics, UAV type, and other relevant attributes.
    \item Implementing memory modules or recurrent architectures to leverage long-term dependencies in acoustic streams.
    \item Exploring Transformer-based backbones and self-supervised pre-training to better utilize large-scale unlabeled acoustic data.
    \item Developing semi-supervised pipelines to rapidly adapt models to new frontline locations where labeled data is scarce.
\end{itemize}

\section*{Acknowledgment}

We are deeply grateful to Zvook \cite{zvookwebsite} for providing the dataset and computational resources essential for this research.

We are especially grateful to the Armed Forces of Ukraine - without their resilience and protection, this work would not have been possible.

This work is part of the master's research conducted by Vadym Vilhurin under the supervision of Volodymyr Sydorskyi at the Institute for Applied System Analysis, Department of Artificial Intelligence, National Technical University of Ukraine "Igor Sikorsky Kyiv Polytechnic Institute."

AI-assisted tools (ChatGPT and Gemini) were used exclusively to improve the clarity and grammar of this text; they did not contribute to the research design, experiments, or analysis.

\bibliographystyle{IEEEtran}
\bibliography{references}

\end{document}